\documentclass[aps,prl,twocolumn,preprintnumbers,amsmath,amssymb,
superscriptaddress,floatfix,a4paper]{revtex4}%

\usepackage{graphicx}
\usepackage{dcolumn}
\usepackage{amsmath}
\usepackage{amssymb}
\usepackage{color}
\usepackage{multirow}
\usepackage{url}

\usepackage[utf8]{inputenc}
\usepackage{siunitx} 
\usepackage{verbatim} 
\usepackage[normalem]{ulem}

\usepackage{placeins} 
\usepackage{float} 

\usepackage{pdfpages}

\usepackage[hidelinks]{hyperref}
\hypersetup{
colorlinks=true,    
linkcolor=blue,
citecolor=blue,
urlcolor=blue,
pdfborder={0 0 0},
bookmarksopen=true,
bookmarksopenlevel=2,
pdfpagemode=UseOutlines,
pdfstartview={Fit},
pdftitle={Manuscript 2H-MoTe2},
pdfauthor={J. A. Krieger},
pdfsubject={2H-MoTe2},
pdfkeywords={soft X-ray ARPES,muon spin rotation,density-functional 
theory, transition-metal dichalcogenide},
pdfhighlight= /I
}

\newcommand{\tHMoTet}{2\textit{H}\protect\nobreakdash-{MoTe}$_2$}

\newcommand{\mSR}{$\mu^+$SR}

\newcommand{\PSI}{Paul Scherrer Institute}

\newcommand{\Tenmr}{${}^{125}$Te}

\newcommand{\scg}{superconducting}

\newcommand{\refsubfig}[2]{\hyperref[#1]{\ref*{#1}#2}}

\begin{document}

\let\oldaddcontentsline\addcontentsline
\renewcommand{\addcontentsline}[3]{}

\title {Hydrogen-impurity induced unconventional magnetism in semiconducting molybdenum ditelluride}

\author{Jonas~A.~Krieger}
\altaffiliation[Current address: ]{Max Planck 
Institut f\"ur 
Mikrostrukturphysik, 
Weinberg 2, 06120 Halle, Germany}
\affiliation{Laboratory for Muon Spin Spectroscopy, Paul Scherrer
Institute, CH-5232 Villigen PSI, Switzerland}
\affiliation{Swiss Light Source, 
Paul Scherrer Institute, CH-5232 Villigen
PSI, Switzerland}
\affiliation{Laboratorium f\"ur Festk\"orperphysik,  ETH Z\"urich, CH-8093
Z\"urich, Switzerland}
\author{Daniel~Tay}
\affiliation{Laboratorium f\"ur Festk\"orperphysik,  ETH Z\"urich, CH-8093
Z\"urich, Switzerland}
\author{Igor~P.~Rusinov}
\affiliation{Tomsk State University, pr. Lenina 36, 634050 Tomsk, Russia}
\affiliation{St. Petersburg State University, Universitetskaya nab. 7/9, 
199034 
St. Petersburg, Russia}
\author{Sourabh~Barua}
\altaffiliation[Current address: ]{Department of Physics, Birla Institute of  
Technology, Ranchi 835215 Jharkhand, India}
\affiliation{Department of Physics, University of Warwick,
Coventry CV4 7AL, UK}
\author{Pabitra~K.~Biswas}
\affiliation{ISIS Facility, Rutherford Appleton Laboratory, Chilton, Didcot, 
Oxon OX110QX, United Kingdom}

\author{Lukas~Korosec}
\affiliation{Laboratorium f\"ur Festk\"orperphysik,  ETH Z\"urich, CH-8093
Z\"urich, Switzerland}

\author{Thomas~Prokscha}
\affiliation{Laboratory for Muon Spin Spectroscopy, Paul Scherrer
Institute, CH-5232 Villigen PSI, Switzerland}
\author{Thorsten~Schmitt}
\affiliation{Swiss Light Source, Paul Scherrer Institute, CH-5232 Villigen
PSI, Switzerland}
\author{Niels~B.~M.~Schr\"oter}
\altaffiliation[Current address: ]{Max Planck Institut f\"ur 
Mikrostrukturphysik, Weinberg 2, 06120 Halle, Germany}
\affiliation{Swiss Light Source, Paul Scherrer Institute, CH-5232 Villigen
PSI, Switzerland}
\author{Tian~Shang}
\altaffiliation[Current address: ]{
Key Laboratory of Polar Materials and Devices (MOE), School of Physics and Electronic Science, East China Normal University, Shanghai 200241, China}
\affiliation{Laboratory for Multiscale Materials Experiments, Paul 
Scherrer Institut,CH-5232 Villigen PSI, Switzerland}
\author{Toni~Shiroka}
\affiliation{Laboratory for Muon Spin Spectroscopy, Paul Scherrer
Institute, CH-5232 Villigen PSI, Switzerland}
\affiliation{Laboratorium f\"ur Festk\"orperphysik,  ETH Z\"urich, CH-8093
Z\"urich, Switzerland}
\author{Andreas~Suter}
\affiliation{Laboratory for Muon Spin Spectroscopy, Paul Scherrer
Institute, CH-5232 Villigen PSI, Switzerland}

\author{Geetha~Balakrishnan}
\affiliation{Department of Physics, University of Warwick,
Coventry CV4 7AL, UK}
\author{Evgueni~V.~Chulkov}
\affiliation{Donostia International Physics Center, P. Manuel de Lardizabal
4, San Sebasti\'an, 20018 Basque Country, Spain}
\affiliation{Departamento de F\'{\i}sica de Materiales UPV/EHU, Centro de
F\'{\i}sica de Materiales CFM - MPC and Centro Mixto CSIC-UPV/EHU, 20080
San Sebasti\'an/Donostia, Spain}
\affiliation{Tomsk State University, pr. Lenina 36, 634050 Tomsk, Russia}
\affiliation{St. Petersburg State University, Universitetskaya nab. 7/9, 
199034 St. Petersburg, Russia}
\author{Vladimir~N.~Strocov}
\email{vladimir.strocov@psi.ch}
\affiliation{Swiss Light Source, Paul Scherrer Institute, CH-5232 Villigen
PSI, Switzerland}
\author{Zaher~Salman}
\email{zaher.salman@psi.ch}
\affiliation{Laboratory for Muon Spin Spectroscopy, Paul Scherrer
Institute, CH-5232 Villigen PSI, Switzerland}

\begin{abstract}
 Layered transition-metal dichalcogenides are proposed as building
 blocks for van der Waals (vdW) heterostructures due to their
 graphene-like two dimensional structure. For this purpose, a
 magnetic semiconductor could represent an invaluable component for
 various spintronics and topotronics devices. Here,
 we combine different local magnetic probe spectroscopies
 with angle-resolved photoemission and density-functional theory
 calculations to show that \tHMoTet\ is 
on the verge of becoming magnetic. Our results present clear evidence that the magnetism can be ``switched on" by a 
hydrogen-like impurity. We also show that this magnetic state
 survives up to the free surface region, demonstrating the material's potential
 applicability as a magnetic component for thin-film heterostructures.
\end{abstract}

\maketitle

\noindent Layered transition-metal dichalcogenides (TMDs) are a
promising class of materials featuring interesting optoelectronic,
\scg, catalytic, and topological properties. They can also be
exfoliated down to monolayers, making them suitable for potential
applications in complex van der Waals (vdW)
heterostructures~\cite{Geim2013,Rajamathi2017}. Among the TMDs,
polymorphic MoTe$_2$, exhibits a variety of electronic and magnetic
properties: Its metallic 1\textit{T}' phase turns into a  \textit{T}$_d$-phase at
$\approx\SI{250}{K}$~\cite{Clarke1978}. The latter is a topological
Weyl semimetal that features unconventional superconductivity at low
temperature~\cite{Sun2015,Qi2016,Nan2018}. 
The 2\textit{H}-phase, the most stable one at ambient conditions,
is an indirect band gap semiconductor
and has been successfully integrated in transistors and
photo-detectors~\cite{Lezama2014,Xu2015,Ding2018}. Moreover, it has
been demonstrated that the 2\textit{H}-phase can be distorted into the
1\textit{T}'-phase as a function of strain or gate
voltage~\cite{Song2016,Wang2017}. In addition, room temperature (RT) magnetic
hysteresis has been observed for small amounts of vanadium deposited
on 2\textit{H}-MoTe$_2$~\cite{Coelho2019}. However, most interestingly,
\tHMoTet\ was shown by muon spin spectroscopy (\mSR) to be a magnetic
semiconductor, exhibiting a long-range magnetic order in the bulk at low
temperatures, which was tentatively attributed to Mo anti-site
defects~\cite{Guguchia2018}. 
However, no other technique has reported such long 
range magnetic order and the origin of this magnetism remains still unclear. In fact, 
theoretical calculations also predict that, apart from anti-site defects, 
hydrogen and transition-metal dopants are able to induce spin polarization in 
MoTe$_2$ in the monolayer limit~\cite{Ma2011,Kanoun2018}. Since a muon can be 
considered to be a light hydrogen isotope, one possible explanation is that the 
magnetism observed in \mSR\ is not intrinsic, but rather induced by the muon 
itself. Another important consideration is that in order to exploit these 
magnetic properties, e.g.~in thin-film heterostructures, one should first elucidate 
their origin and evolution as a function of depth in the vicinity of an 
interface (with vacuum or other materials).

Here, we address these questions by utilizing complementary local spin probe 
measurements, with implanted muons,  as well as with the host's 
\Tenmr-NMR to reveal the origin of the reported magnetism. We find that while some of 
the implanted muons, which form a hydrogen-like state in \tHMoTet, detect 
magnetism in this material, \Tenmr-NMR shows no evidence of any  
intrinsic magnetic order. These findings were then confirmed by detailed 
density-functional theory (DFT) calculations, reproducing the experimental 3D 
electronic band structure. When including isolated hydrogen (i.e., muon like)  impurities, 
the DFT unveils an induced magnetic moment at the impurity.  Furthermore, we show that this 
hydrogen-induced magnetic state remains unchanged towards the crystal surface, 
establishing \tHMoTet\ as a switchable magnetic semiconductor by hydrogen 
doping, thus paving the path towards new 
possibilities for integrating this intriguing TMD into future applications in 
thin-film heterostructures.

\begin{figure*}[tb]
\includegraphics[width=0.95\linewidth]{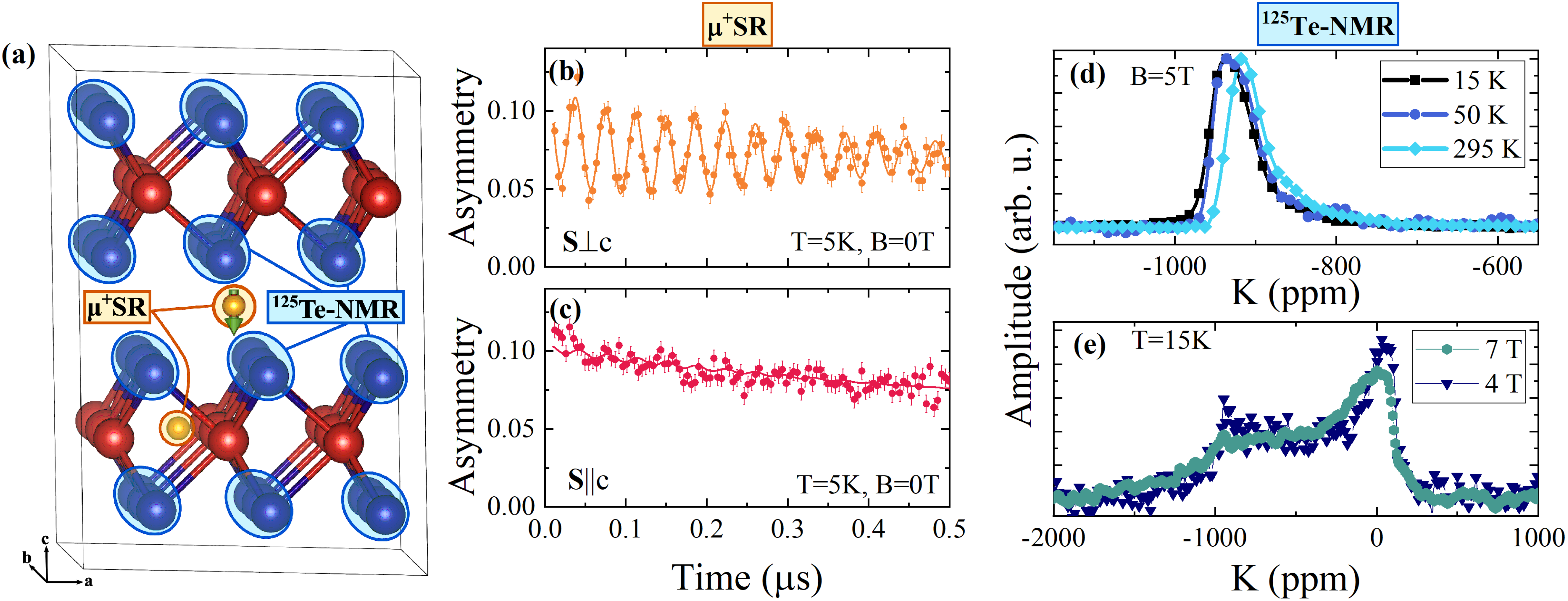}
  \caption{ 
  \textbf{(a)} Schematic illustration of the 
different probing sites. 
The muon's  stopping sites are interstitial, inside the layer and in the vdW gap, 
whereas \Tenmr-NMR is sensitive to the local magnetism at the Te site. The  
\mSR\ asymmetry spectra measured  in zero field at $T=$\SI{5.5}{K} with the initial muon spin polarization \textbf{(b)} perpendicular ($\perp$) and \textbf{(c)}  parallel  ($\parallel$) to the $c$-axis of the crystal. Coherent zero-field precession is observed only in \textbf{(b)}, indicating that the muons probe a 
magnetic field along the $c$-axis.
\textbf{(d)}
\Tenmr-NMR spectral line shapes of a \tHMoTet\ single 
crystal in a field of 
$\SI{5.01}{T}$ perpendicular to 
the $c$-axis at different temperatures. \textbf{(e)} \Tenmr-NMR powder line 
shapes at \SI{15}{K} at different applied fields. The lack of a pronounced 
temperature or field dependence in the \Tenmr-NMR lines implies the absence of an intrinsic 
magnetic order.
}
  \label{fig:LocalProbe}
\end{figure*}

In order to gain a better understanding of the nature of the previously 
observed magnetic signal in \tHMoTet, we combine different local magnetic 
spectroscopies, (i)~\mSR\ which uses individually implanted, spin-polarized muons, 
stopping at interstitial sites in the lattice, and (ii)~\Tenmr-NMR on the 
tellurium atoms, intrinsic to the host compound. This is illustrated in 
Fig.~\refsubfig{fig:LocalProbe}{(a)}.
Both the muon and the \Tenmr\ nuclei have a spin of $1/2$ and are, 
therefore, unaffected 
by quadrupolar interactions and thus act as purely magnetic probes of the 
system.
The \tHMoTet\ single crystals used in this study were grown by the Chemical Vapour Transport technique, using TeCl$_4$ as the transporting agent. 
In contrast, the previously studied samples from Ref.~\cite{Guguchia2018} were 
prepared using Te flux. Therefore, we expect  the density and nature of native 
defects in our samples to be different. This subtle difference should affect their magnetic 
properties, if   long-range magnetism were induced by 
structural defects. 

Indeed we observe a coherent precession of the muon spin in zero 
magnetic field and at low temperature, see 
Fig.~\refsubfig{fig:LocalProbe}{(b)}. However, our \mSR\ measurements show identical
results  to 
those published earlier, namely yielding the same transition
temperature and the same internal magnetic field magnitude. 
The precession frequency at \SI{5.5}{K} in Fig.~\refsubfig{fig:LocalProbe}{(b)} 
corresponds to an internal magnetic field of 
\SI{199.1(3)}{mT}, consistent with the previously reported value of 
$\mu_0H_{\mathrm{int}}=\SI{200}{mT}$~\cite{Guguchia2018}. 
Additional measurements as a function of 
temperature and using low-energy muons stopping within \SIrange{10}{100}{nm} from 
the surface are presented in Supplementary Fig.~S8 and show a similar behavior~\cite{SI}.
Therefore, the local magnetic
properties observed with \mSR\ in \tHMoTet\ are robust and do not
change in the near-surface region of the crystals, nor do they depend
on the details of the sample synthesis. This is an indication that the 
previously reported structural defects~\cite{Guguchia2018} are unlikely to be the 
origin of long-range magnetic order.
 
Furthermore, measurements as a function of the angle between the $c$-axis and initial muon spin polarization, as shown in  Fig.~\refsubfig{fig:LocalProbe}{(b,c)}, 
reveal that the internal magnetic field sensed by the muon is oriented along the $c$-axis. 
A maximal Larmor precession amplitude is detected when the initial muon spin 
$\mathbf{S}\perp c$-axis (Fig.~\refsubfig{fig:LocalProbe}{(b)}) while it 
almost vanishes for $\mathbf{S}\parallel c$-axis 
(Fig.~\refsubfig{fig:LocalProbe}{(c)}). This is a clear indication that the 
local magnetic field sensed by the implanted muons points along the 
crystallographic $c$-axis. A detailed analysis of the full 
angular dependence between the initial muon spin and the $c$-axis is presented in 
Supplementary Fig.~S6 and supports this 
conclusion~\cite{SI}.

Contrary to the \mSR\ results, we find no signatures of 
intrinsic magnetism when we investigate \tHMoTet\ with \Tenmr-NMR. 
Example 
spectra  on single 
crystalline \tHMoTet\ in a field of $\approx\SI{5.01}{T}$ are shown in 
Fig.~\refsubfig{fig:LocalProbe}{(d)}.  The spectra 
neither exhibit a significant broadening nor a resonance shift as a function of 
temperature, which  is strong evidence that the \Tenmr\ nuclear spin experiences no significant changes in 
the local spin susceptibility. This is in stark contrast to the typical NMR 
hallmarks of magnetic ordering, which consist of a 
temperature-dependent NMR line shift and broadening, caused by the emergence of 
an internal magnetic field near the magnetic transition, as observed for 
example in  unenriched Fe~\cite{Budnick1961} or in the RAlGe material 
family~\cite{Daniel2020}. Similar distinctive NMR magnetic signatures are also 
expected for defect-induced 
magnetism, as has been observed e.g.~in the 2D system SiC~\cite{Zhang2017}. 

Furthermore, we exclude the presence of a static magnetic field at the 
\Tenmr\ site, by measuring powder line shapes at \SI{15}{K} in different 
applied fields, see Fig.~\refsubfig{fig:LocalProbe}{(e)}. These spectra are 
significantly broader than those of single crystals, a broadening which we attribute 
to an anisotropic chemical shift. However, apart from a small change in the 
line shape, the line width scales 
perfectly with the applied field. This lack of field dependence is indicative 
of 
the non-magnetic origin of the line broadening, as it excludes the presence of 
a static internal magnetic field at the \Tenmr\ site.
Therefore, these results clearly show that an intrinsic spin probe 
exhibits no 
evidence of any long-range magnetic order in \tHMoTet. This is in clear contrast to our 
observation with \mSR\ and a strong indication that the local static fields 
observed with muons originate from a muon-induced effect.

\begin{figure*}[tb]
  \includegraphics[width=0.85\linewidth]{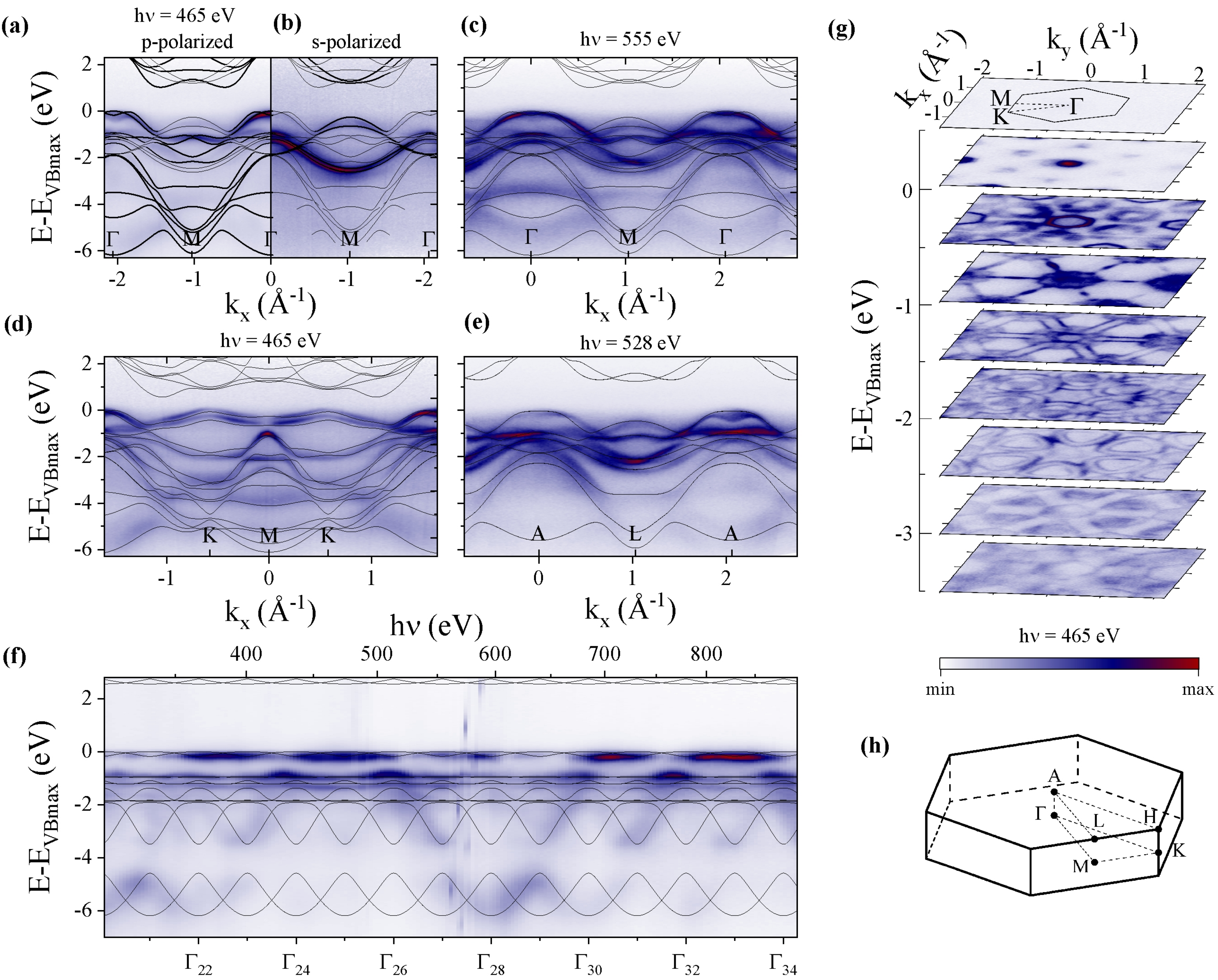}
    \caption{
      Comparison of  soft X-ray ARPES spectra at
      $\sim$\SI{12}{K} with the calculated band structure (solid
      lines). \textbf{(a,b)} Dispersion along $\Gamma-M$ at
      $\Gamma_{25}$, \textbf{(c)} along $\Gamma-M$ at $\Gamma_{27}$,
      \textbf{(d)} along K-M at $\Gamma_{25}$, \textbf{(e)} along A-L
      between $\Gamma_{26}$ and $\Gamma_{27}$ and \textbf{(f)} along
      $\Gamma$-A.  \textbf{(g)} Constant energy cuts at $\Gamma_{25}$.
      In \textbf{(a)} and \textbf{(b)} symmetric and antisymmetric
      bands were probed by using $p$- and $s$- polarized light,
      respectively. There, the width of the calculated lines
      represents the contribution of symmetric or anti-symmetric
      orbitals to the band structure. All other spectra were probed
      with circularly polarized light, which does not distinguish between
      symmetric and anti-symmetric states.  \textbf{(h)} Brillouin
      zone and high symmetry points of \tHMoTet.  }
\label{fig:ARPES}
\end{figure*}

In order to reveal the origin of this effect, we use 
DFT to calculate the ground state of \tHMoTet\ and how it is affected by an 
individual implanted muon. We start by refining our DFT calculation to reproduce the electronic structure of \tHMoTet. For this purpose, we 
determine the three dimensional electronic band dispersion with soft X-ray 
angle-resolved photoemission spectroscopy (ARPES). Spectra along high symmetry
directions at photon energies corresponding to the $\Gamma$ and A
planes are shown in Fig.~\refsubfig{fig:ARPES}{(a-e)}.
These reveal sharp bands over the whole range of binding energies. We
find that the valence band maximum is formed at the $\Gamma$ point by
a band derived from states that are symmetric along $\Gamma-M$. In
addition, we measure the band dispersions in $k_z$ along $\Gamma-A$ as
a function of photon energy (Fig.~\refsubfig{fig:ARPES}{(f)}). Along
this direction we expect that a two dimensional state forms a flat band
with no clear dispersion. In contrast, the dispersive spectral weight
in Fig.~\refsubfig{fig:ARPES}{(f)} reveals that most bands hybridize
across the vdW gap and gain a strong 3D character, despite the 2D
layered structure of \tHMoTet. Interestingly, the spectral weight
exhibits modulations with longer periodicity in $k_z$ than the period of the
reciprocal lattice. Such an effect has been previously observed in
other 2\textit{H}-TMDs and is connected with the
non-symmorphic symmetry group of the 2\textit{H} structure of MoTe$_2$~\cite{Weber2018}.

We compare the spectral weight in Fig.~\ref{fig:ARPES} to the
calculated band structure. Note that in general, ab-initio DFT of vdW
materials with strong spin-orbit coupling can be misleading, as
exemplified by {\textit{T}$_d$-MoTe$_2$}, where an on-site correlation term ($U$)
had to be introduced, to correctly reproduce the experimental
results~\cite{Rhodes2017,Nan2018,Aryal2019}. However, our DFT
calculations, performed without a $U$ term, are in excellent agreement
with the experimental bands. This indicates that, unlike
{\textit{T}$_d$-MoTe$_2$}, in \tHMoTet\ there are only weak correlations. We further note 
that there are no signatures of intrinsic 
magnetism in the calculations of bulk \tHMoTet.

\begin{figure*}[tb]
  \includegraphics[width=0.99\linewidth]{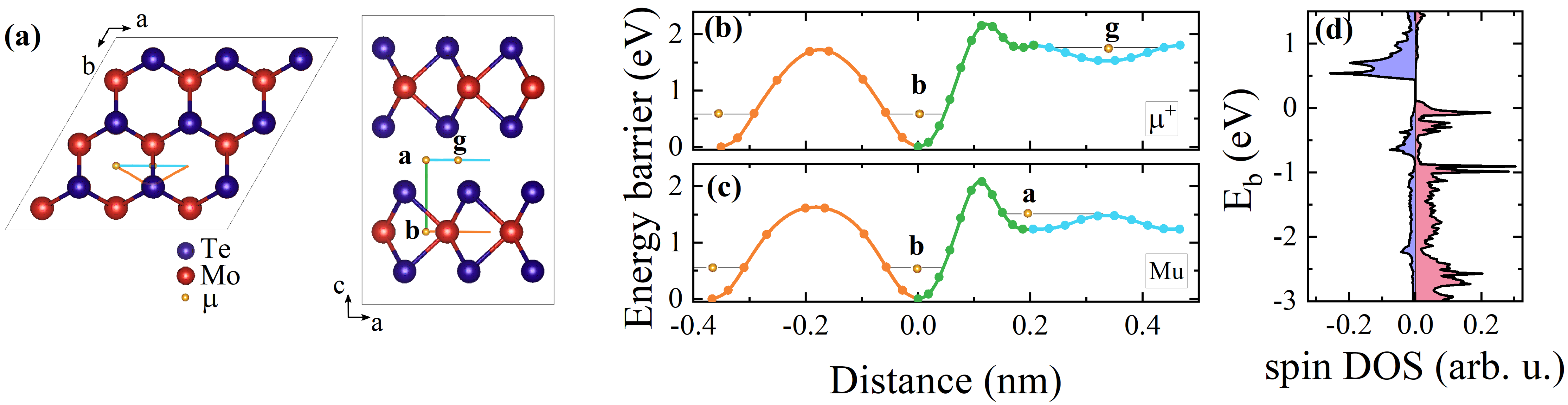}
  \caption{ 
  \textbf{(a)} Stopping sites and diffusion paths indicated
    in the 3$\times$3$\times$1 supercell.
    The muon ($\mu^{+}$) at sites 
    \textbf{b}~and~\textbf{g} and 
     muonium (Mu) at sites \textbf{b}~and~\textbf{a}.
    Energy barrier for \textbf{(b)} $\mu^{+}$, and \textbf{(c)} Mu
    diffusion along the different paths. The 
    black lines show
    the ground-state energy level of the particles at the local
    minima.  
    \textbf{(d)}
    Calculated local spin density of states for a muon 
stopping at site \textbf{a}, assuming a nonzero Hubbard $U$.}
\label{fig:Diffusion}
\end{figure*}

By correctly reproducing the 3D band dispersions, we have identified a
reliable representation of \tHMoTet's electronic structure and charge
density within DFT. We can now employ it to predict the behavior of
implanted  $\mu^+$ spin probes, introduced as isolated
impurities at interstitial positions within a supercell.
We consider two charge states of the
muon by adding a charge on the supercell ($+1$ for a
diamagnetic $\mu^+$ and $+0$ for a neutral muon state)~\cite{Moeller2013,Bernardini2013}. The electrically 
neutral state of a muon corresponds to a muonium (Mu), a hydrogen-like bound 
state between a muon and an electron.
Figure~\ref{fig:Diffusion} shows the resulting stable stopping sites
and the diffusion energy barrier along the most energetically
favorable paths between them.
In particular, we find that the diamagnetic $\mu^+$ has a stable site \textbf{b} (denoted by its Wyckoff letter) inside the Mo-layer and a 
meta-stable site \textbf{g} in the vdW gap. 
On the other hand, the neutral muonium state can be stable in \textbf{b} or 
meta-stable in \textbf{a}.  

Finally, we consider Mu- and muon-induced magnetism in these sites. Indeed, for the Mu at the 
site~\textbf{a}, and when employing a finite Hubbard term $U$ on the hydrogen (i.e.~Mu)  potential, a significant spin 
density is induced in the material, Fig.~\refsubfig{fig:Diffusion}{(d)}, while all other sites remain nonmagnetic (Supplementary Fig.~S5)~\cite{SI}. Such muon-induced magnetism is very 
similar to what was found 
for hydrogen adsorbed on a monolayer of \tHMoTet\ or on 
graphene~\cite{Ma2011,GonzalezHerrero2016}.
As we discuss in more detail in the Supplementary~\cite{SI}, such a magnetic muonium state is 
consistent with both the zero-field as well as the transverse field behavior (Supplementary Fig.~S7) of 
the observed \mSR\ signals.
Therefore, we conclude that the \tHMoTet\ does not exhibit an intrinsic magnetic 
order and that the magnetic signal observed with \mSR\ originates from a magnetic 
muonium state in site \textbf{a}. As the muon can be considered a light 
hydrogen 
isotope, this induced magnetic state should be the same for a hydrogen atom located at 
site \textbf{a}.

To conclude, using local magnetic probe measurements, ARPES and 
theoretical 
DFT calculations, we are able to reveal that \tHMoTet\ is a semiconductor on the 
verge of becoming magnetic. The comparison of the intrinsic \Tenmr-NMR probe 
with the  implanted  muon spin probe confirms that the 
observed local field in \mSR\ does not reflect an intrinsic long-range magnetic 
order. Instead, we find that the muonium, being a hydrogen-like isolated 
impurity, can be used to ``switch-on'' a local magnetic moment in this system. 

Importantly, we find that the effect of muonium on the local magnetic
properties 
remains unchanged up to the surface of the crystals. This paves the path towards 
the use of hydrogen impurities to transform \tHMoTet\ into a magnetic semiconductor 
building block for vdW heterostructures. We expect that  the use 
of such a material is not limited to applications in spintronics 
devices~\cite{Baltz2018} but may also include, e.g.,
Majorana heterostructures~\cite{Sau2010} or introduction of magnetic
proximity at the surface of topological materials~\cite{Hesjedal2017}.

\section*{Acknowledgements}
This work is partially based on experiments performed at the Swiss
Muon Source (S$\mu$S) and Swiss Light Source (SLS), Paul Scherrer
Institute, Villigen, Switzerland.  
J.A.K.~acknowledges
support by the Swiss National Science Foundation (SNF-Grant
No.~200021\_165910). 
I.P.R. acknowledges support from the Ministry of Education and Science of the
Russian Federation within State Task Nr.~0721-2020-0033.
The work at the University of Warwick was supported by EPSRC, UK, through through Grants EP/M028771/1 and EP/T005963/1.
The magnetization measurements were supported by the Swiss National Science 
Foundation (SNF-Grant No.~206021\_139082).
We are grateful to C.~W.~Schneider for help with single crystal XRD measurements.
We thank Z. Guguchia, and S. Holenstein from PSI, CH, and A.~Chatzichristos, D.~Fujimoto, V.~L.~Karner, R.~M.~L.~McFadden, 
J.~O.~Ticknor, W.~A.~MacFarlane, and R.~F.~Kiefl from the University of British 
Columbia, CA, for helpful discussions.
Further, we note that a preliminary version of these results was published as 
part of a PhD thesis, Ref.~\cite{KriegerThesis}.


\onecolumngrid
\pagebreak
\clearpage
\begin{center}
\section*{\large \textit{Supplemental Material:} Hydrogen-impurity induced unconventional magnetism in semiconducting molybdenum ditelluride}
\end{center}
\setcounter{equation}{0}
\newcounter{FiguresInMainText}
\setcounter{FiguresInMainText}{\value{figure}}
\setcounter{table}{0}
\makeatletter
\renewcommand{\theequation}{S\arabic{equation}}
\renewcommand{\thefigure}{S\the\numexpr\value{figure}-\value{FiguresInMainText}}
\renewcommand{\bibnumfmt}[1]{[S#1]}
\renewcommand{\citenumfont}[1]{S#1}

\tableofcontents
\let\addcontentsline\oldaddcontentsline

\section*{Sample growth and characterization}
\begin{figure}[hbt]
  \includegraphics[width=0.50\linewidth]{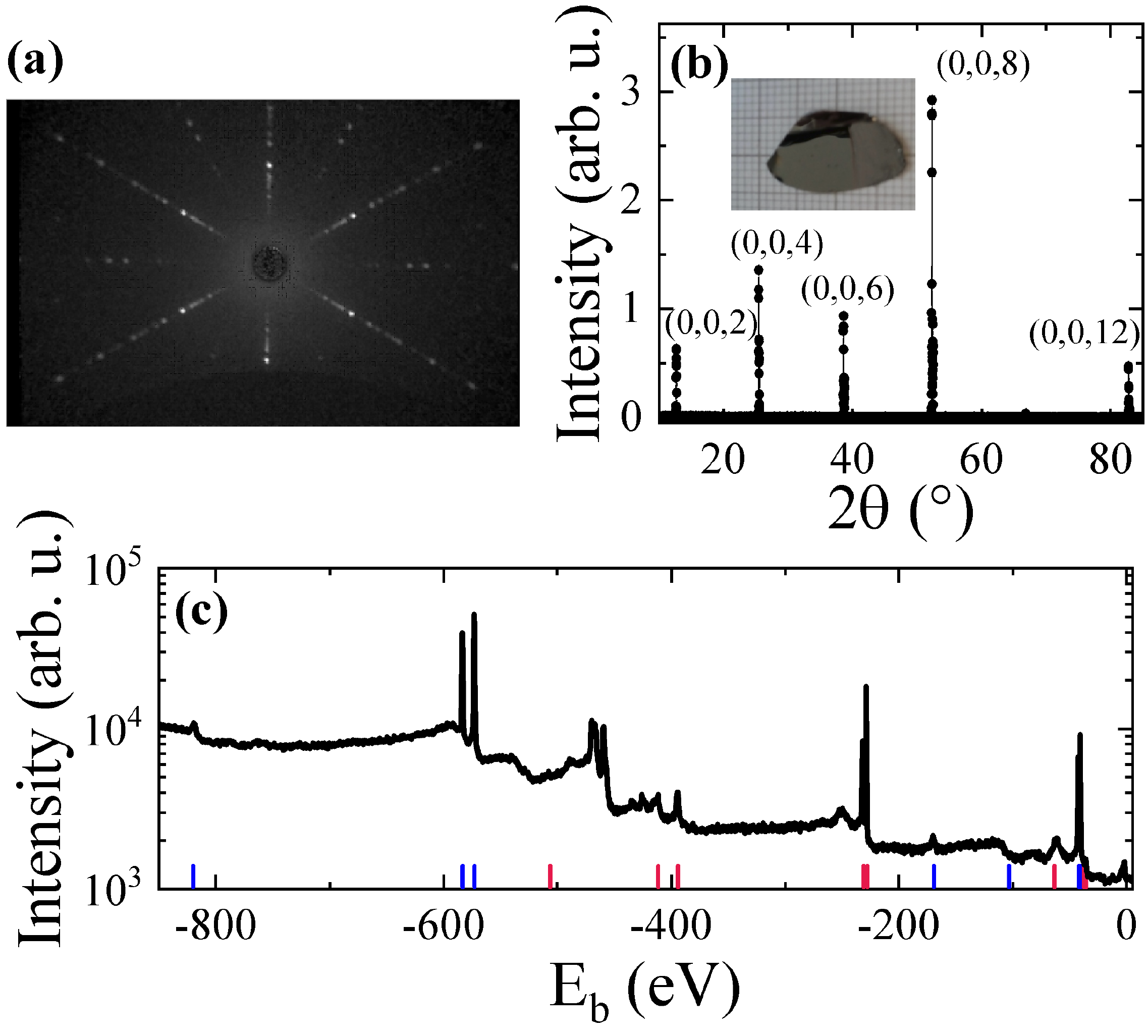}
  \caption{\textbf{(a)} The Laue back reflection X-ray diffraction pattern for 
one of the single crystals. 
   \textbf{(b)} X-ray (0,0,n) diffraction peaks. The inset shows a photograph 
of one of the single crystals.
  \textbf{(c)} XPS core level spectrum measured at $h\nu=\SI{950}{eV}$. The 
blue (red) lines show the expected positions of the Mo (Te) peaks. 
  }
\label{fig:prep}
\end{figure}

High-quality single crystals of \tHMoTet\ were grown by chemical vapor
transport method. Stoichiometric amounts of the starting materials, Mo
and Te, along with the transport agent TeCl$_4$
($\SI{3}{mg/cm^{3}}$), were mixed together in a quartz ampule and
sealed in vacuum. The quartz tube was then placed for 3~weeks in a two-zone
horizontal tube furnace with the charge end at \SI{800}{\celsius} and
the end where the crystals form at \SI{750}{\celsius}.
The high quality of the single crystals and the crystal structure was
confirmed by x-ray diffraction (XRD) and x-ray photoelectron spectroscopy (XPS), as shown in Figure~\ref{fig:prep}. The additional XPS peaks  
around \SI{-460}{eV} are the Te~{M$_{4,5}$\nobreakdash-N$_{4,5}$N$_{4,5}$} Auger 
lines. 

\subsection*{Magnetization measurements}

\begin{figure}[hbt]
  \includegraphics[width=0.99\linewidth]{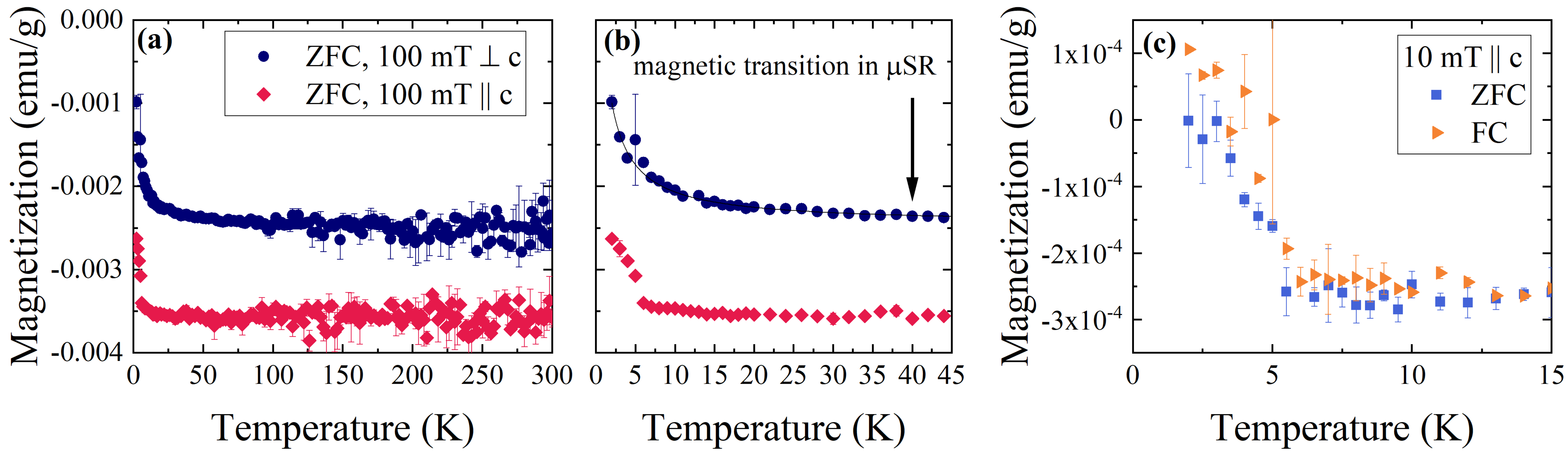}
  \caption{ Magnetization measurements on \tHMoTet. \textbf{(a)} Zero field cooled (ZFC) 
magnetization in an applied 
field of \SI{100}{mT}. \textbf{(b)} Zoom into the low temperature region of 
\textbf{(a)}. The solid line shows the fit of a Curie-Weiss law 
($T_c=\SI{-0.9(5)}{K}$). The arrow shows the onset of the magnetic zero field 
oscillations observed in \mSR. \textbf{(c)} Field cooled (below \SI{50}{K}) and 
zero 
field 
cooled magnetization for a \SI{10}{mT} out-of-plane field. 
  }
\label{fig:SQUID}
\end{figure}

Magnetometry measurements have been performed with a Quantum Design 
superconducting quantum interference device (SQUID) magnetic properties 
measurement system (MPMS) and are shown in Fig.~\ref{fig:SQUID}. 
For an 
applied in-plane field of \SI{100}{mT}, the sample shows a 
{Curie\nobreakdash-Weiss} like behavior. In contrast, if the field is applied 
out-of-plane 
there is a signature of a transition around \SI{5}{K}, see 
Fig.~\refsubfig{fig:SQUID}{(b)}. Moreover, for an out-of-plane field of 
\SI{10}{mT} this transition features  a small hysteresis in the 
magnetization signal, Fig.~\refsubfig{fig:SQUID}{(c)}. 
Its magnitude is too small for an intrinsically ferromagnetic sample and it cannot account for the observed magnetic transition in \mSR, which appears at much higher temperatures ($\sim\SI{40}{K}$).  
However, it could be attributed to various origins, including magnetic 
defects~\cite{SIGuguchia2018} or 
magnetic grain-boundaries~\cite{SITongay2012}.

\section*{ARPES}
The soft X-ray ARPES experiments were performed on the X03MA beamline at the Swiss Light Source 
of the \PSI, Villigen, Switzerland~\cite{SIStrocov2010,SIStrocov2014}. The
analyzer slit was oriented along the incoming beam direction and the
photon energy $h\nu$ was tuned in the range of
\SI{320}{eV}~to~\SI{900}{eV}. The high photon energies allow a better
$k_z$ resolution~\cite{SIStrocov2012} than in previous investigations of
\tHMoTet~\cite{SIBoker2001}.  The combined beamline and analyzer
resolution was better than~\SI{0.15}{eV} at $h\nu=$\SI{900}{eV} and 
better than \SI{63}{meV} around $h\nu\sim$\SI{500}{eV}.  The samples
were cleaved in-situ and kept at $\sim\SI{12}{K}$ in a vacuum
lower than \SI{e-10}{mbar}.  We only present the binding energy with
respect to the valence band maximum, because \tHMoTet\ tends to
statically charge upon x-ray illumination, which causes a rigid shift
of the spectrum, see below.

\subsection*{Out-of plane dispersion}
In the soft X-rays energy range the photoelectrons have a larger escape 
depth compared to standard ultraviolet ARPES. This reduces the $k_z$ 
broadening and allows to determine the out-of plane dispersion. For a measured 
photoelectron kinetic energy $E_{\mathrm{kin}}$, the perpendicular momentum is 
given by:
\begin{equation}
\mathbf{k}_{z}=\sqrt{\frac{2m_{\mathrm{e}}\left(E_{\mathrm{kin}}-V_{000}
\right) } { \hbar^2 }-\mathbf{k}_{\parallel} }+\mathbf{p}_{\gamma,z},
\end{equation}
where $m_{\mathrm{e}}$ is the electron mass, $\mathbf{k}_{\parallel}$ the 
in-plane momentum, $V_{000}$ the inner potential and 
$\mathbf{p}_{\gamma,z}$ the component of the photon momentum ($h\nu/c$) 
perpendicular to the sample surface~\footnote{The formulas for 
the specific measurement geometry can be found in Ref.~\cite{SIStrocov2014}}. The 
inner potential is a material and energy dependent parameter. We have estimated 
it by comparing the measured band structure with the expected periodicity of 
the Brillouin zone (cf.~Fig.~1(f) in the main paper). We  used a value of 
$V_{000}\approx\SI{10}{eV}$, which gives a fair agreement between theory and 
experiment, but slightly differs from the previously used 
$V_{000}=\SI{16(1)}{eV}$~\cite{SIBoker2001} .

\subsection*{Static charging}
\begin{figure}[hbt]
  \includegraphics[width=0.5\linewidth]{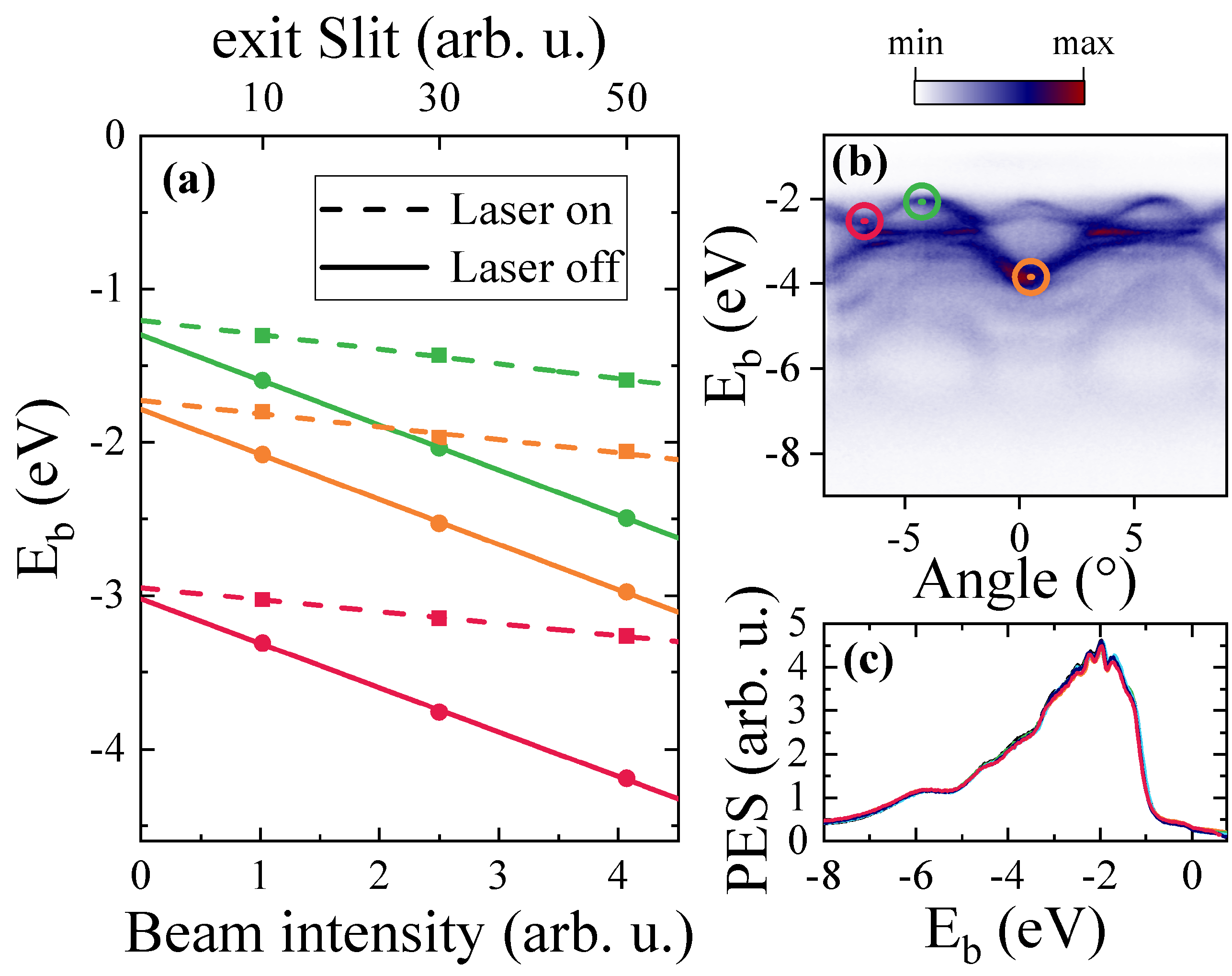}
  \caption{\textbf{(a)} Binding energy of several features indicated in 
   \textbf{(b)} as a function of x-ray flux. 
  \textbf{(c)} Angle integrated spectra corresponding to \textbf{(a)}, after 
rigidly aligning the valence band maximum.
  }
\label{fig:charging}
\end{figure}

Because \tHMoTet\ is a semiconductor, it tends to statically charge upon x-ray 
irradiation at low temperature. This is shown in Fig.~\ref{fig:charging}, which 
displays the binding energy of several features in the band structure as a 
function of photon flux. 
The binding energies were calibrated by measuring the 
Fermi level of Au.
For some measurements the sample was additionally irradiated with a 
near~UV~laser. This excites some charge carriers and helps to significantly 
reduce the charging (Fig~\refsubfig{fig:charging}{(a)}). In both cases, the 
charging can be assumed to be a rigid shift of the spectrum. This is evident 
from the linear dependence on the flux in Fig~\refsubfig{fig:charging}{(a)} and 
from the collapse of angle integrated spectra after alignment of the valence 
band maximum in Fig~\refsubfig{fig:charging}{(c)}. Therefore, in the main paper 
we only show the binding energy with respect to the valence band maximum.

\subsection*{n-type samples}
\begin{figure}[hbt]
  \includegraphics[width=0.99\linewidth]{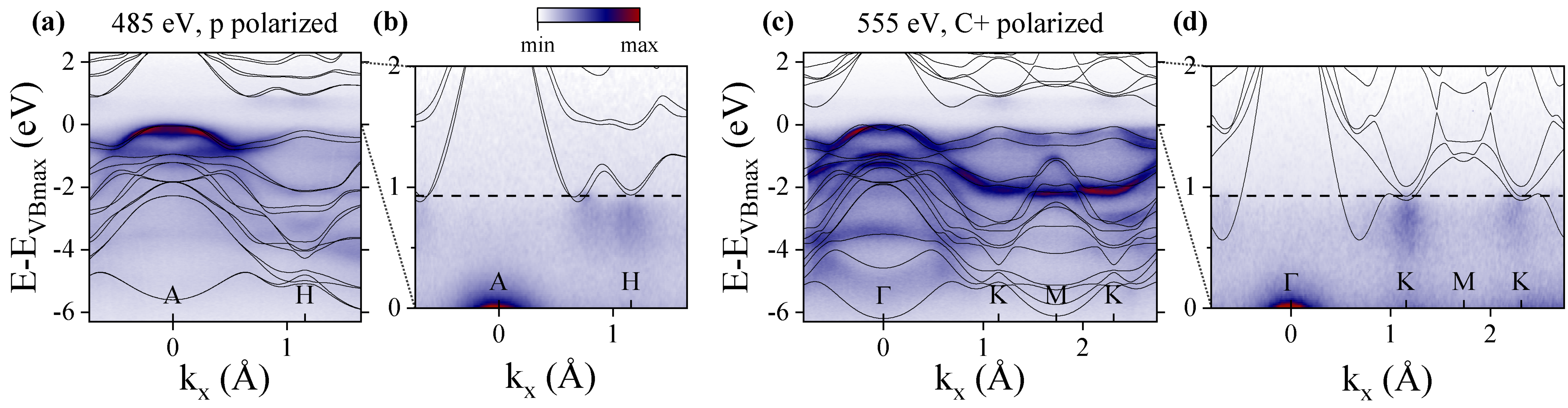}
  \caption{\textbf{(a,c)} Spectra measured on a sample that became n-type over 
time. \textbf{(b,d)} Zooms into the region above the valence band maximum. The 
dashed line indicates the estimated, effective position of the Fermi level.
  }
\label{fig:nType}
\end{figure}

All measurements shown in the main text were performed on samples that 
were freshly cleaved in-situ at $\sim$\SI{12}{K} directly before the 
measurement. However, after a couple of hours of ARPES measurements the 
\tHMoTet\ samples usually 
degrade and the material becomes slightly $n$-type. This is exemplified by 
the spectra in Fig.~\ref{fig:nType}, where a small spectral weight above the valence band was observed. 
These measurements clearly confirm that the band gap in \tHMoTet\ is indirect.

\section*{Density-functional theory calculations}

The electronic structure was calculated with Density-Functional Theory (DFT) and the Projected Augmented Wave (PAW)
method from the VASP code~\cite{SIKresse1996,SIKresse1999}. Spin-orbit
coupling was treated within the scalar-relativistic approach by the
second variation method. The structure was relaxed and the 
exchange-correlation functional was
chosen to be Perdew-Burke-Ernzerhof (PBE)~\cite{SIPBE}.
The parameter for the relaxed cell used in the ARPES calculation were given by 
$a=\SI{3.491}{\angstrom}$, $c=\SI{13.647}{\angstrom}$, where the Mo and Te sit 
at the $2c$ and $4f$ Wyckoff positions of space group 194, respectively, with 
relative 
$z$-coordinates $0.116$ and $0.383$ for Te.
Along $\Gamma$-M an 
orbital decomposition into symmetric 
($p_z$,$p_y$,$d_{z^2}$,$d_{x^2-y^2}$,$d_{zy}$) and antisymmetric 
($p_x$,$d_{xy}$,$d_{xz}$) orbitals was preformed. The orbital character does 
not fully separate into symmetric and antisymmetric bands because of spin-orbit 
coupling.
Nevertheless, when using $p-$ ($s-$) polarized light 
[Fig.~2(a)~(2(b)) in the main text], only the 
symmetric (antisymmetric) part of the electronic wave function, which primarily 
consists of these orbitals, will be probed.

\subsection*{Muon site determination}
In order to
simulate a single muon impurity, we used a hydrogen
pseudopotential in the relaxed $3\times 3 \times 1$ supercell of 2\textit{H}-MoTe2. The assumed lattice constants in the $3\times 3\times 1$ supercell, 
c$=\SI{13.647}{\angstrom}$ and a corresponding a$=\SI{10.4745}{\angstrom}$, were the same as for the band structure calculations. 
The charge state of the muon is accounted for by assuming a neutral/charged 
supercell~\cite{SIMoeller2013,SIBernardini2013}. 
Therein, we preformed nudged elastic band (NEB)~\cite{SIHenkelman2000} as well 
as density of states calculations 
 on the basis of the Linear
Combination of Pseudo Atomic Orbitals (LCPAO) approach as implemented
in the OPENMX code~\cite{SIPAO}. 
The muon ground state energies were estimated by approximating the 
NEB diffusion energy
barriers with an anisotropic harmonic oscillator potential. The crystal 
structures were visualized with \texttt{VESTA}~\cite{SIVESTA}. 
Some further details regarding the stable muon positions are listed in 
Tab.~\ref{tab:StableSites}.
\begin{table}[ht]
    \centering
    \begin{tabular}{cclllcc}\hline\hline
     Species&Wyckoff position&\multicolumn{3}{c}{frac. 
coordinates}&$\Delta E$~$(eV)$& ZPE~$(eV)$\\ \hline
     $\mu^+$&$2b$&0&~~0~~~&0.25&0&0.57\\
     $\mu^+$&$6g$&0.5&~~0~~~&0.5&1.5&0.22\\\hline
     Mu&$2b$&0&~~0~~~&0.25&0&0.54\\
     Mu&$2a$&0&~~0~~~&0.5&1.24&0.26\\\hline
     \hline\hline
     \end{tabular}
     \caption{Details of the identified stable and 
meta-stable site for several implanted probe species. $\Delta E$ denotes the 
difference in the total energy of the relaxed supercell compared to the most 
stable site and ZPE is the approximate ground state energy of the muon. 
}\label{tab:StableSites}
\end{table}

\begin{figure}[hbt]
  \includegraphics[width=0.95\linewidth]{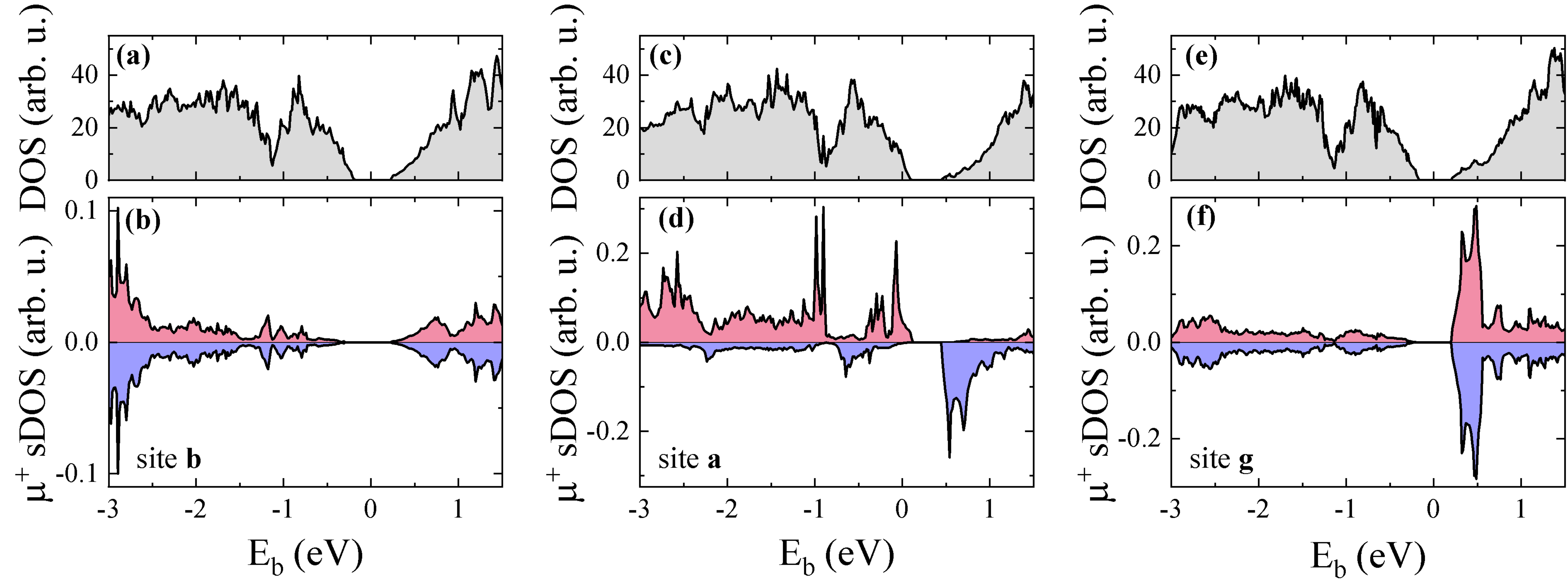}
  \caption{Total density of states and local spin density of states calculated 
for a muon  (i.e.~a hydrogen potential in a charged supercell) at 
site~\textbf{b} 
\textbf{(a,b)}, site~\textbf{a} \textbf{(c,d)}, and site~\textbf{g} 
\textbf{(e,f)}, 
respectively. 
Note that site~\textbf{a} is not a stable stopping site for a diamagnetic muon. 
  }
\label{fig:sDOS}
\end{figure}

We further show the local spin density of states at the different 
stopping sites in Fig.~\ref{fig:sDOS}.
It is evident from this figure, that there is no induced magnetism at the 
stable 
muon stopping site~\textbf{b}~and~\textbf{g}, 
see~Fig.~\refsubfig{fig:sDOS}{(b,f)}.
For site~\textbf{b}, the 
absence of an induced spin polarization can be understood because of  
the strong  pairwise interaction of the muon with the neighboring Mo atoms. 
However, in the case of a muonium (and also a muon) sitting at the hexagonal 
site 
inside the vdW gap (site~\textbf{a}) and when employing a
  finite Hubbard term $U$, a significant spin density is induced,
see~Fig.~\refsubfig{fig:sDOS}{(c,d)}. This is very similar to what was found 
for hydrogen adsorbed on a monolayer  
\tHMoTet\ or on graphene~\cite{SIMa2011,SIGonzalezHerrero2016}.
Note that site~\textbf{a} is only a stable stopping site of muonium but not of 
a muon. 
As argued in more detail in the main text and below, such a muonium induced magnetic state can 
explain the observed coherent muon spin precession, while being consistent with 
the absence of a magnetic signal in magnetization and \Tenmr-NMR.

\section*{Muon spin spectroscopy}
The \mSR\ experiments were performed on the general purpose spectrometer 
(GPS) at the Swiss Muon Source, \PSI,
Switzerland~\cite{SIAmato2017}. The temperature was controlled using a $^4$He 
flow cryostat. In order to vary the orientation of the 
sample with respect to the initial muon spin direction, the sample was mounted on a mechanical 
rotation-stage.
 Nearly $\SI{100}{\percent}$ spin
polarized muons were implanted. The asymmetric emission of the muon's decay
positrons (lifetime $\tau_\mu=\SI{2.2}{\micro s}$) enables us to use them  as a 
local magnetic
probe~\cite{SIYaouanc2011}. 
Note, that at any 
given moment, there is only one muon in the sample. Therefore, we can exclude 
that a muon probes magnetism induced by a nearby Mu or $\mu^+$. 
The data were analyzed
with \texttt{musrfit} software suite~\cite{SImusrfit}.

\subsection*{Zero field angle dependence}
\begin{figure}[hbt]
  \includegraphics[width=0.45\linewidth]{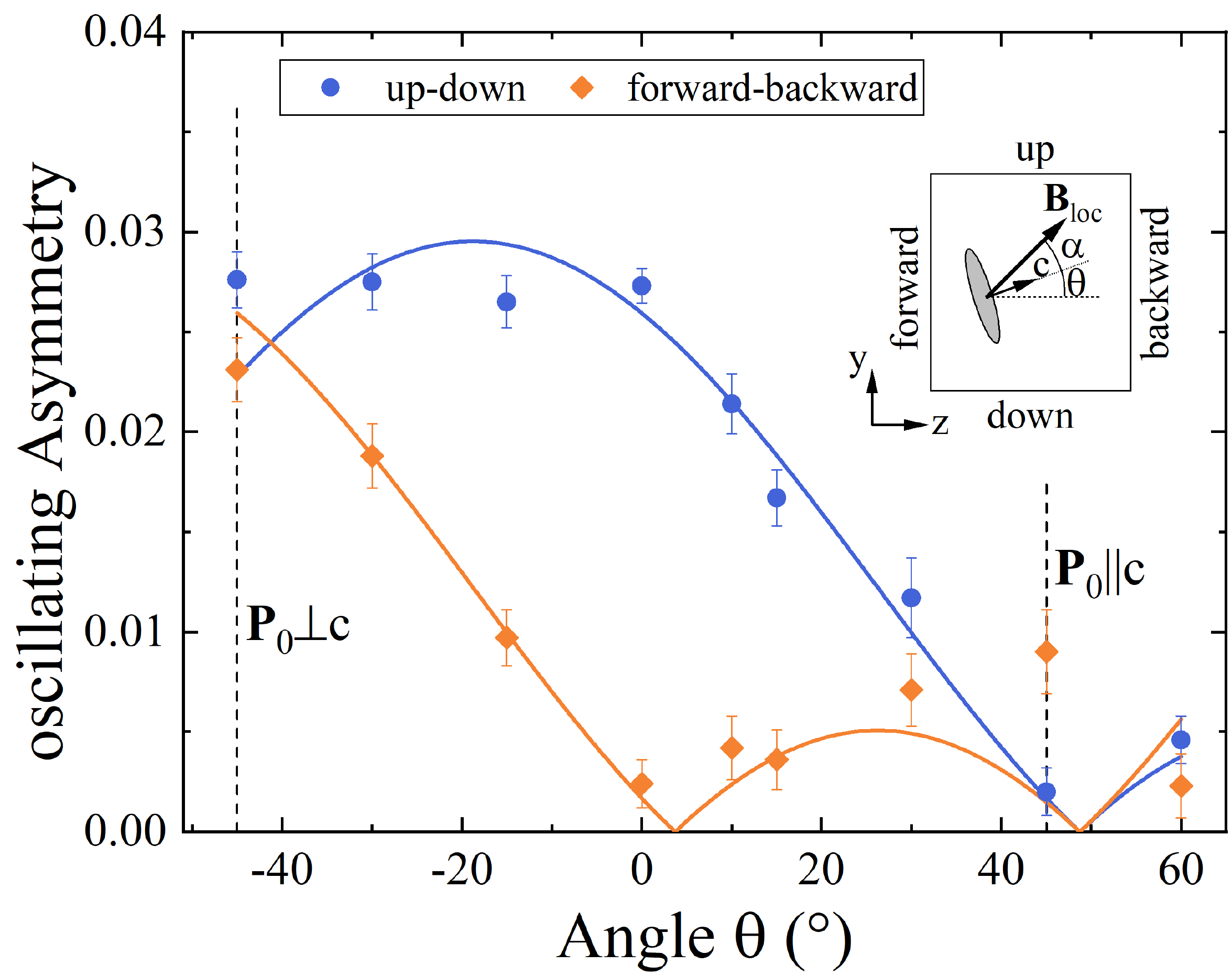}
  \caption{Oscillating asymmetry in different sets of detectors, measured for 
different orientations of the sample. The inset depicts the 
measurement geometry.
  }
\label{fig:rotationGPS}
\end{figure}

In order to determine the variation in the zero field (ZF) \mSR\ spectra as a 
function of 
sample orientation, the asymmetry spectra (such as the ones shown in 
Fig.~1(b,c) of 
the main paper) were fitted to a sum of a Gaussian damped oscillating part and 
two exponentially damped components:
\begin{equation}
 A(t)=A_1e^{-\frac{1}{2}\left({\sigma}
t\right)^2}\cos\left(\gamma_\mu B 
t+\varphi\right)+A_2e^{-\lambda_{\mathrm{slow}}t}+A_3e^{-\lambda_{\mathrm{fast}}
t}.
\end{equation}
We further assumed that the internal field $B$ and the depolarization rate 
$\sigma$ of the oscillating part to be independent of the measured projection of the polarization.
Figure~\ref{fig:rotationGPS} shows the resulting oscillating ZF asymmetry 
$A_1$ in 
different sets of detectors (corresponding to different projections of the muon spin polarization) as a function of the sample orientation. 
In order to understand this rotation dependence, 
one has to consider the Larmor precession of the muon polarization 
$\mathbf{P}(t)$, defined as 
\begin{equation}\label{eq:LarmorDgl}
 \frac{\rm{d}{\mathbf{P}}(t)}{\rm d t}=\gamma_\mu \mathbf{P}(t)\wedge 
\mathbf{B}.
\end{equation}
Here, $\mathbf{B}$ denotes the local magnetic field vector and $\gamma_\mu 
= 2 \pi \times \SI{135.5}{MHz/T}$ is the gyromagnetic ratio of the muon.
The resulting precession can be written 
as~\cite{SIHerak2013,SIYaouanc2011}:
\begin{align}\label{eq:LarmorTime}
 \mathbf{P}(t)&=\mathbf{p}^\parallel+ 
\mathbf{p}^\perp_1\cos(\gamma_\mu Bt)+
\mathbf{p}^\perp_2\sin(\gamma_\mu Bt)
 ,\\
 \mathbf{p}^\parallel&=\frac{(\mathbf{P}_0\cdot 
\mathbf{B})\mathbf{B}}{B^2}
,
\qquad
 \mathbf{p}^\perp_1=\mathbf{P}_0-\mathbf{p}^\parallel
 ,
 \qquad
 \mathbf{p}^\perp_2=\mathbf{p}^\perp_1\wedge\frac{\mathbf{B}}{B},
\end{align}
where $\mathbf{P}_0$ is initial polarization vector and $B=\|\mathbf{B}\|$ denotes 
the magnitude of the field.  

We choose the coordinate system $(x,y,z)$ depicted in the inset of 
Fig.~\ref{fig:rotationGPS}, with ${\mathbf{P}_0\propto(0,1,1)}$, and the local 
field $\mathbf{B}=B(0,\sin{(\theta+\alpha)},\cos{(\theta+\alpha)})$. The angle 
$\theta$ denotes the rotation of the sample's $c$-axis with respect to the 
detectors and $\alpha$ is a hypothetical offset between the local magnetic 
field 
seen by the muon and the $c$-axis of the crystal. 

The asymmetry probed by a set of opposite detectors is proportional to the 
projection of the polarization along that direction. Therefore, in the inset of 
Fig.~\ref{fig:rotationGPS} the asymmetry in the up-down detectors is 
$A_{\text{up-do}}\propto \mathbf{P}(t)\cdot \hat{\mathbf{e}}_y$, and the one in 
the forward backward detectors 
$A_{\text{fw-ba}}\propto \mathbf{P}(t)\cdot \hat{\mathbf{e}}_z$, where 
$\hat{\mathbf{e}}_y$ and $\hat{\mathbf{e}}_z$ are the unit vectors in the $y$ 
and $z$ directions.

Inserting these conventions into Eq.~\ref{eq:LarmorTime}, we find that the 
oscillating part of the asymmetry is given by $\mathbf{p}^\perp_1$ and that 
accordingly
\begin{align}
 A_{\text{up-do}}&\propto1-\sqrt{2}\sin{(\alpha+\theta + 
\pi/4)}\sin{(\theta+\alpha)},\\
 A_{\text{fw-ba}}&\propto1-\sqrt{2}\sin{(\alpha+\theta + 
\pi/4)}\cos{(\theta+\alpha)}.
\end{align}
These curves are depicted as solid lines in Fig.~\ref{fig:rotationGPS}, where 
$\alpha=-\SI{4(1)}{\degree}$ was determined using nonlinear least-squares 
optimization.
As the sample was aligned with respect to the detector frame by hand, this 
small value of $\alpha$ could easily correspond to a small 
misalignment of the sample. Therefore, we conclude that within our measurement 
precision, the local field points along the $c$-axis of the crystal.

\subsection*{Field dependence in transverse field}
\begin{figure}[hbt]
  \includegraphics[width=0.85\linewidth]{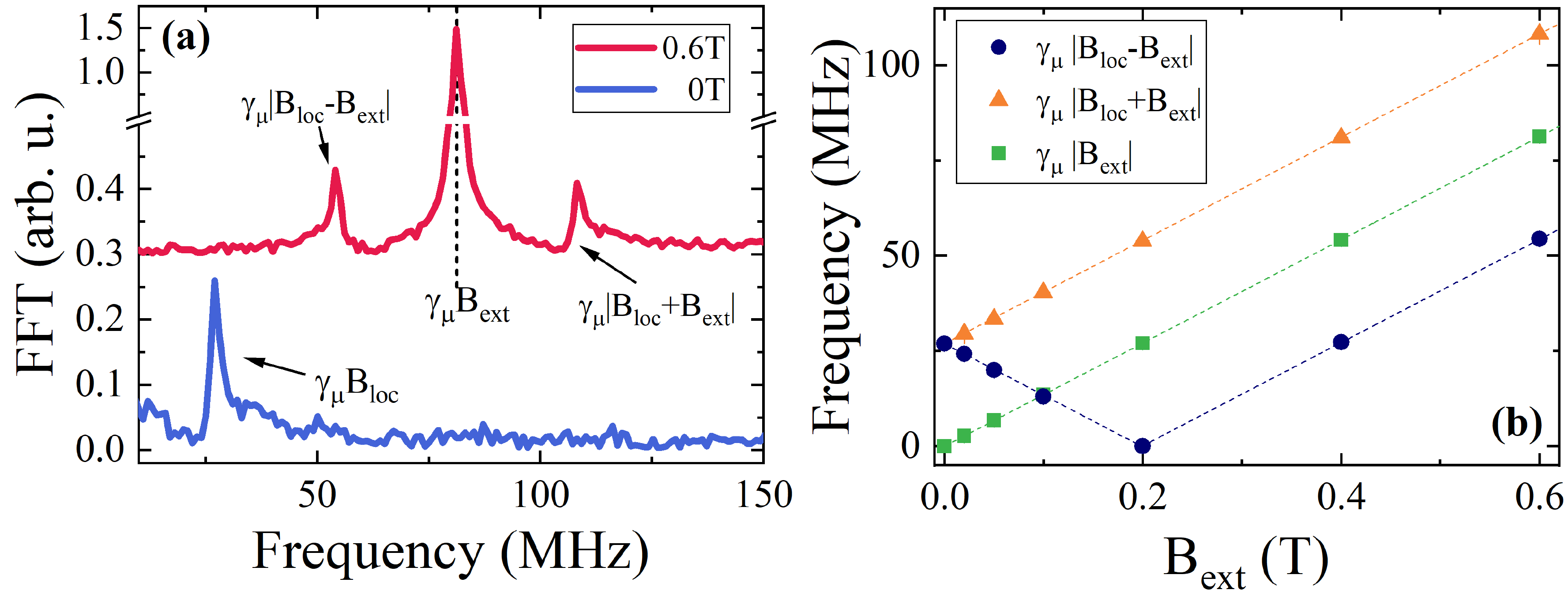}
  \caption{\textbf{(a)} Fourier transform of the \mSR\ asymmetry spectra measured in different applied fields. The curves have been offset for clarity. \textbf{(b)} 
Evolution of the oscillation frequencies as a function of the applied field.
  }
\label{fig:BscanGPS}
\end{figure}

We have also determined the evolution of the muon oscillation frequencies  as 
a function of a transverse field applied along the crystallographic $c$-axis. The 
fast Fourier transforms of representative spectra are shown in 
Fig.~\refsubfig{fig:BscanGPS}{(a)}.
We find that there is one component in the spectra, 
which oscillates at a frequency corresponding to the applied field, and which 
allows an accurate calibration of applied field ($B_{\rm ext}$). This component is partially attributed to a background contribution from a silver plate, onto 
which 
the sample was mounted. Other contributions to this component could also include some 
diamagnetic muons, which experience no strong magnetic field in \tHMoTet. In addition, we observe other oscillating components which split up into peaks at 
$\gamma_\mu|B_{loc}\pm B_{ext}|$, where 
$B_{loc}\sim\SI{200}{mT}$ is the field corresponding to the ZF oscillation.
The accurate values of these frequencies were extracted by fitting the \mSR\ 
time spectra to a sum of Gaussian damped oscillating components plus a 
non-oscillating Gaussian damped contribution:
\begin{equation}
 A(t)=A_0e^{-\frac{1}{2}\left({\sigma_0}t\right)^2}+
\sum_n{A_ne^{-\frac{1}{2}\left({\sigma_n}
t\right)^2}\cos\left(\gamma_\mu B_n 
t+\varphi\right)}
.
\end{equation}
The resulting field values $B_n$ are shown in
Fig.~\refsubfig{fig:BscanGPS}{(b)}.
We note that the internal field $B_{loc}$ seems to point equally likely along 
the positive and negative $c$-axis direction.
At a first glance, this seems incompatible with a muonium state, to which we 
attribute this signal in the main part of the paper. Usually the behavior of 
muonium in a field is dominated by the hyperfine coupling and described by the 
Hamiltonian
\begin{equation}\label{eq:MuHamiltonian}
\mathcal{H}=\hbar\mathbf{I}_\mu A\mathbf{S}_e+\hbar\gamma_e\mathbf{S}_e
\cdot\mathbf{B}_{\mathrm{ext}}+\hbar\gamma_\mu\mathbf{I}_\mu\cdot\mathbf{B}_{
\mathrm {ext}},
\end{equation}
where $A$ is the hyperfine coupling tensor, $\mathbf{I}_\mu$, and 
$\mathbf{S}_e$ are the spins of the muon and the bound electron and 
{$\gamma_e= 2\pi\times\SI{28}{GHz/T}$} is the electron's 
gyromagnetic ratio. 
In a system without locally induced magnetism, this 
results in the presence of four different oscillation frequencies, which scale 
non-linearly as a function of field, see 
e.g.~Refs.~\cite{SIPatterson1988,SICox2003,SIYaouanc2011}.

However, in the case of muonium induced magnetism, we can assume that the 
electron spin $\mathbf{S}_e$ is locked along its local easy axis, and remains 
unaffected by the (relatively small) applied fields. Then 
Eq.~\ref{eq:MuHamiltonian} can be simplified to,
\begin{equation}\label{eq:magMuHam}
\mathcal{H}=\hbar\mathbf{I}_\mu 
A\mathbf{S}_e+\hbar\gamma_\mu\mathbf{I}_\mu\cdot\mathbf{B}_{
\mathrm {ext}}=
\hbar\gamma_\mu\mathbf{I}_\mu\cdot(\mathbf{B}_{loc}+\mathbf{B}_{
\mathrm {ext}}),
\end{equation}
where $A\mathbf{S}_e/\gamma_\mu=:\mathbf{B}_{loc}$ is an effective 
local magnetic field characterizing the ZF oscillation. The field dependence of 
the oscillation frequencies resulting from Eq.~\ref{eq:magMuHam} is fully 
consistent with our observation presented in Figs.~\ref{fig:rotationGPS},~\ref{fig:BscanGPS}, with $\mathbf{B}_{loc}=(\pm\SI{200}{mT})\mathbf{\hat{c}}$, where $\mathbf{\hat{c}}$ is a unit vector along the $c$-axis.

\subsection*{Low-energy $\mu$SR}

\begin{figure*}[hbt]
  \includegraphics[width=0.99\linewidth]{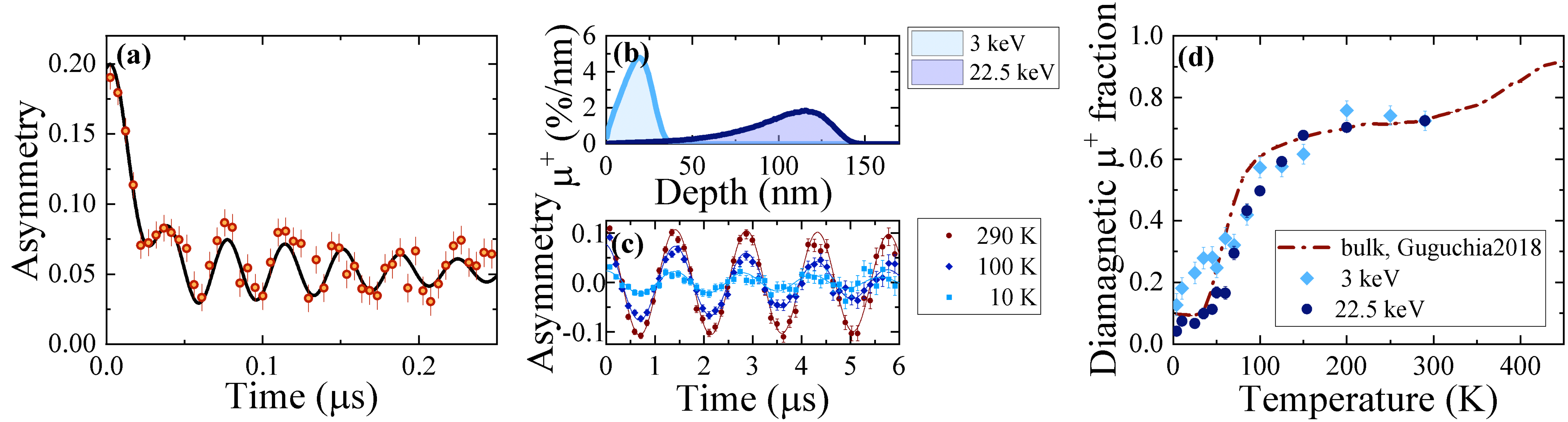}
  \caption{\textbf{(a)} \mSR\ time spectrum measured in zero field at
    \SI{10}{K} with an implantation energy of \SI{22.5}{keV}.
    \textbf{(b)} Simulated implantation profiles for different
    implantation energies. The average stopping depths beneath the
    surface for \SI{3}{keV}~or~\SI{22.5}{keV} are~\SI{18}{nm}
    and~\SI{99}{nm}, respectively.  \textbf{(c)} Representative weak
    transverse field asymmetry spectra at \SI{22.5}{keV} corresponding
    to \textbf{(d)}. \textbf{(d)} Fraction of diamagnetic muons as a
    function of temperature, normalized at RT.  The dashed line shows
    the bulk data from Ref.~\cite{SIGuguchia2018}.  The solid lines in
    \textbf{(a,c)} show fits to the data.  }
\label{fig:LEM}
\end{figure*}

We also investigate the evolution of the observed 
magnetism within $\sim\SI{100}{nm}$ from the surface using low 
energy \mSR\
at the $\mu$E4
beamline of the Swiss Muon Source, \PSI,
Switzerland~\cite{SIProkscha2008}. 
For these measurements, multiple single crystals were glued on a Ni
coated sample plate with the $c$-axis parallel to the applied field, but
perpendicular to the initial muon spin. The implantation depth was controlled 
by varying the kinetic energy,  $E$,  of the incident muon beam. The resulting stopping 
profiles were  calculated with the \texttt{Trim.SP} software~\cite{SIEckstein1991} and are shown in Fig.~\refsubfig{fig:LEM}{(b)}.
At \SI{10}{K} and
$E=$\SI{22.5}{keV} (corresponding to an average implantation depth of
\SI{99}{nm}), the muon spins exhibit a coherent
precession in zero field (ZF) similar to what we observed in the 
bulk, see Fig.~\refsubfig{fig:LEM}{(a)}.
This spectrum was fitted to the following 
function:
 \begin{equation}\label{eq:muZF}
  A(t)=A_1e^{-\frac{1}{2}\left({\sigma_{\mathrm{osc}}}
t\right)^2}\cos\left(\gamma_\mu B_{\mathrm{int}} 
t\right)+A_2e^{-\lambda_{\mathrm{slow}}t}+A_3e^{-\frac{1}{2}\left({\sigma_{
\mathrm {fast}}}t\right)^2}.
 \end{equation}
The first term describes the muons that oscillate in a local magnetic 
field $B_{\mathrm{int}}$. The second term represents the 
slowly depolarized (nonmagnetic) signal and the third accounts for the strongly 
damped part due to background contribution from muons stopping in the Ni coated sample plate.
This analysis is similar to that used in Ref.~\cite{SIGuguchia2018}, except  for 
the background contribution. We find that the oscillation frequency in 
Fig.~\refsubfig{fig:LEM}{(a)} corresponds to an 
internal static field
$B_{\mathrm{int}}=$\SI{200(2)}{mT}. Within the
statistical error of the measurement, this is consistent with what is 
found in the bulk~\cite{SIGuguchia2018}.

We then use a small magnetic field (\SI{5}{mT}, transverse to the initial muon
spin) to identify the fraction of diamagnetic muons, which is
proportional to the non-magnetic fraction of the sample
[Fig.~\refsubfig{fig:LEM}{(c,d)}]. 
These spectra were fitted with a single, 
Gaussian damped oscillating polarization.
\begin{equation}
 A(t)=A_0e^{-\frac{1}{2}\left(\sigma t\right)^2}\cos\left(\gamma_\mu 
Bt\right).
\end{equation}
This function only describes the muons in a nonmagnetic environment oscillating 
at a frequency close to the Larmor frequency in the applied field. Any magnetic or strongly 
damped signal (cf.~Eq.~\ref{eq:muZF}) leads to an effective loss of $A_0$. 
 The background 
contribution to $A_0$ was determined by measuring an empty Ni plate with the 
same beamline settings. This temperature independent value was subtracted from 
the measured~$A_0$ and the resulting curve is shown in 
Fig.~\refsubfig{fig:LEM}{(d)}.
Independent of depth and
  within the experimental uncertainties, all points in
Fig.~\refsubfig{fig:LEM}{(d)} closely follow the bulk behavior
reported in Ref.~\cite{SIGuguchia2018}. Therefore, we conclude that the bulk magnetic
properties observed with \mSR\ in \tHMoTet\ do not
change in the near-surface region of the crystals.

\section*{\Tenmr-NMR}
The \Tenmr-NMR signals were collected by means of spin-echo detection  in a $\pi/2$--$\pi$ sequence, using a 
\SIrange{62.5}{125}{MHz} duplexer, with a \SI{63}{dB} preamplifier. The line 
shapes were then extracted via a fast Fourier transformation of those signals.
Typical pulse widths of \SI{5}{\micro s} and repetition times  
of \SI{20}{s} were used, with each line shape corresponding to more than 20'000 
records.
Measurements on a single crystal were performed in a field of 
$\SI{5.007}{T}$ perpendicular to
the $c$-axis, while those on powders were performed in fields of \SI{7.067}{T} 
and \SI{4.011}{T}. 
The \Tenmr\ reference frequencies were calibrated using $^{27}$Al-NMR on aluminum 
foil, which is known to have a very small shift as a function of 
temperature~\cite{SIHarris2001}. 

\subsection*{Powder line shapes}
\begin{figure}[hbt]
  \includegraphics[width=0.4\linewidth]{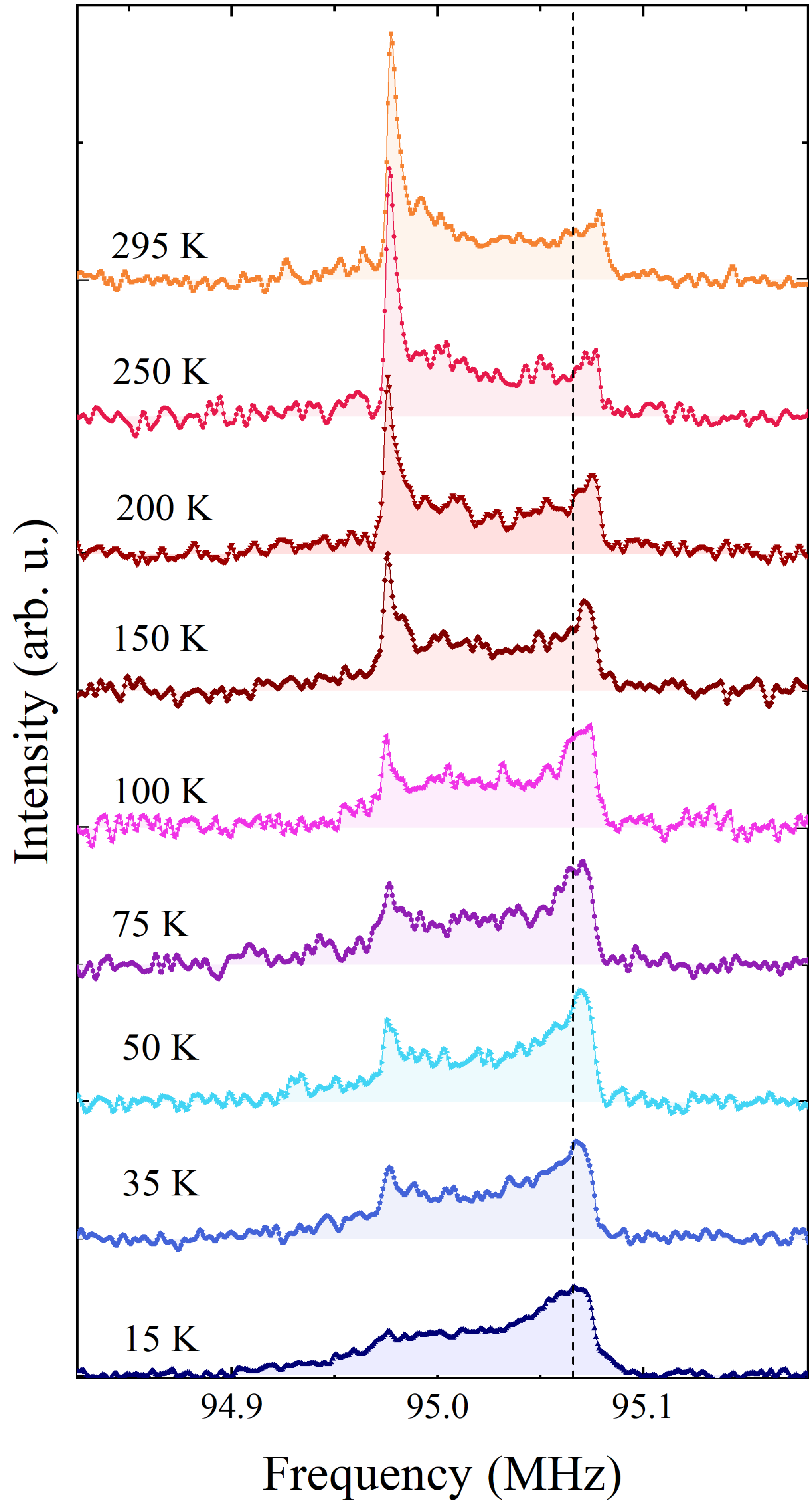}
  \caption{\Tenmr-NMR powder line shapes of \tHMoTet\ in a field of 
\SI{7.067}{T} at different temperatures. The spectra have been offset for 
clarity. The black dashed line indicates the expected position of the reference 
line.
  }
\label{fig:NMRlines}
\end{figure}

Figure~\ref{fig:NMRlines} shows the detailed temperature dependence of the 
powder line shape of \tHMoTet\ in \SI{7.067}{T}. Similar to the spectra of 
the single crystal measurements shown in the main part of the paper, here, too, we observe
no significant peak-shift or broadening, that would be the two common fingerprints of a 
magnetic transition.
We 
note that there are some small changes in the line shape as a 
function of temperature. However, these are most likely due to a change in spectral 
weight between different portions of the resonance line, owing to a difference in their spin-lattice relaxation (SLR) temperature dependence.

\subsection*{Spin-lattice relaxation}
\begin{figure}[hbt]
  \includegraphics[width=0.45\linewidth]{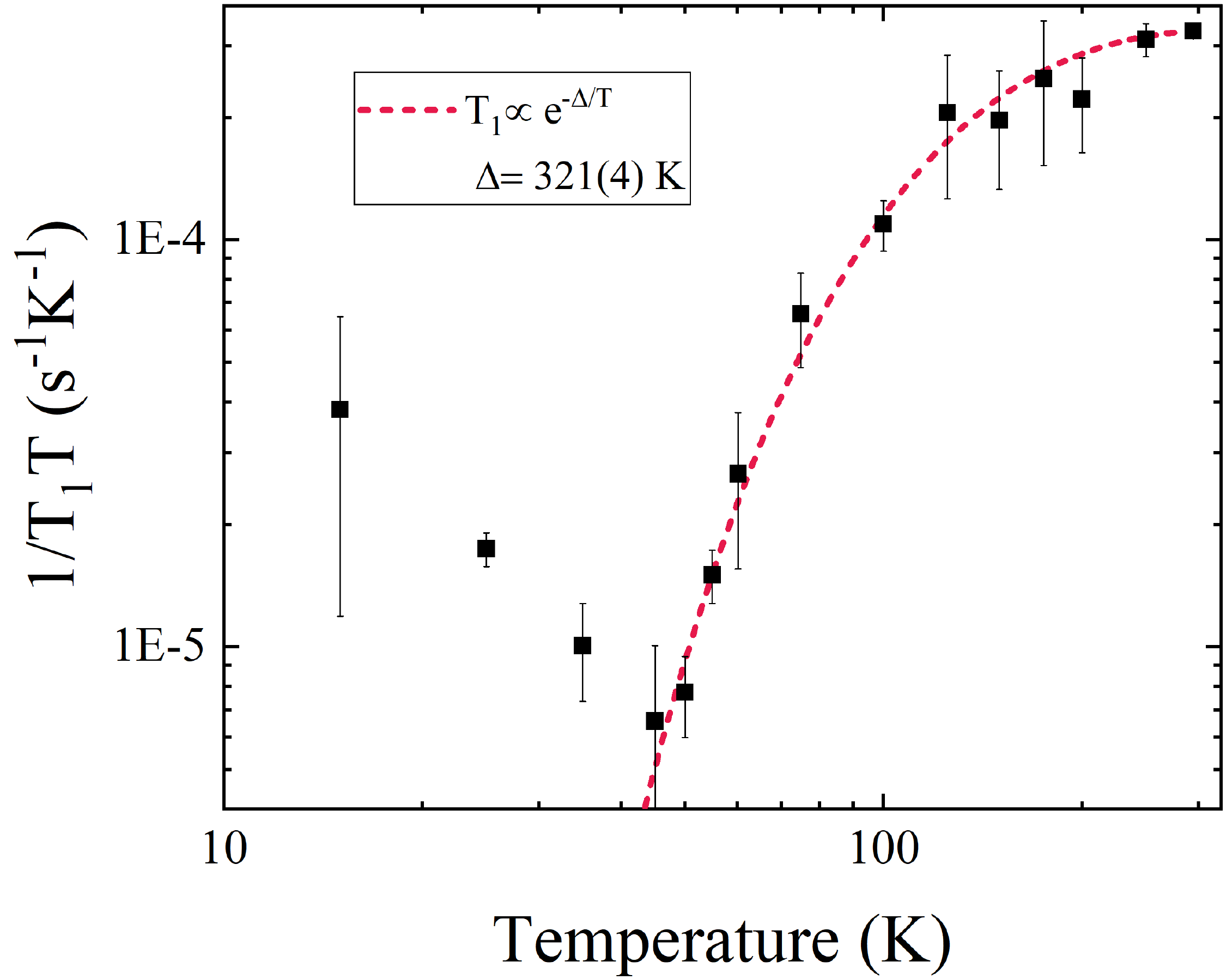}
  \caption{\Tenmr-NMR spin lattice relaxation rate in \tHMoTet\ in a field of 
\SI{7.067}{T} as a function of temperature. The  dashed line indicates the 
activated Arrhenius-like behavior at high temperature.
  }
\label{fig:NMRSLR}
\end{figure}

The spin-lattice relaxation rate ${T_1}^{-1}$ of the \Tenmr\ moments is shown in 
Fig.~\ref{fig:NMRSLR}.
 At high temperature, the SLR shows an exponential 
dependence on temperature ($1/T_1\propto e^{\Delta/T}$). This is a typical 
behavior observed in semiconductors. However, we find that here the gap 
$\Delta=\SI{27.7(3)}{meV}$ is much smaller than the spectroscopic band gap, 
which is on the order of $\sim\SI{1}{eV}$, cf.~Fig.~\ref{fig:nType}. This difference 
is likely due to, e.g., the presence of a small amount of impurity-induced 
in-gap states, which have too little spectral weight to be detected with ARPES, 
but still can depolarize the \Tenmr\ nuclear spins.
At present, we are unable to determine the mechanism for the 
increase in relaxation rate at low temperatures in \tHMoTet. However, we note 
that a similar increase in relaxation was observed in the analogue compounds 
ZrTe$_2$ and ZrSiTe~\cite{SITian2020,SITian2021}. In both cases, the increase in 
relaxation rate is suggested 
to arise from a  contribution from Dirac electrons. However, \tHMoTet\ is not a 
Dirac system, so the fact that there is a similar increase in relaxation rate 
suggests that the true origin of this effect might be due to some new mechanism, 
yet to be understood. 
Finally, we note that the SLR has a minimum around \SI{40}{K}, i.e., the onset temperature of the 
magnetic oscillations observed by \mSR. However, in the presence of an intrinsic 
magnetic transition, the SLR should instead show a peak at this temperature due 
to critical magnetic fluctuations. This finding confirms once more that the observed local magnetic field in \mSR\ is not intrinsic to \tHMoTet.

 \FloatBarrier

\end{document}